\documentclass[preprint]{aastex631}

\usepackage{upgreek}
\usepackage{amsmath,amsthm,amssymb}
\usepackage{mathdots} 
\usepackage{mathrsfs} 
\usepackage{physics}
\usepackage[normalem]{ulem}

\graphicspath{{./images}}
\usepackage{subfigure}

\begin{document}

\title{Energization of charged test particles in magnetohydrodynamic fields: waves vs turbulence picture}


\author{F.Pugliese}
\affiliation{CONICET - Universidad de Buenos Aires, Instituto de Física Interdisciplinaria y Aplicada (INFINA), Ciudad Universitaria, 1428 Buenos Aires, Argentina.}
\affiliation{Universidad de Buenos Aires, Facultad de Ciencias Exactas y Naturales, Departamento de Física, Ciudad Universitaria, 1428 Buenos Aires, Argentina.}

\author{M. Brodiano}
\affiliation{CONICET - Universidad de Buenos Aires, Instituto de Física Interdisciplinaria y Aplicada (INFINA), Ciudad Universitaria, 1428 Buenos Aires, Argentina.}
\affiliation{Universidad de Buenos Aires, Facultad de Ciencias Exactas y Naturales, Departamento de Física, Ciudad Universitaria, 1428 Buenos Aires, Argentina.}

\author{N. Andr\'es}
\affiliation{Universidad de Buenos Aires, Facultad de Ciencias Exactas y Naturales, Departamento de Física, Ciudad Universitaria, 1428 Buenos Aires, Argentina.}
\affiliation{Universidad de Buenos Aires - CONICET, Instituto de Astronomía y Física del Espacio (IAFE), Ciudad Universitaria, 1428, Buenos Aires,Argentina}

\author{P. Dmitruk}
\affiliation{CONICET - Universidad de Buenos Aires, Instituto de Física Interdisciplinaria y Aplicada (INFINA), Ciudad Universitaria, 1428 Buenos Aires, Argentina.}
\affiliation{Universidad de Buenos Aires, Facultad de Ciencias Exactas y Naturales, Departamento de Física, Ciudad Universitaria, 1428 Buenos Aires, Argentina.}

\begin{abstract}
Direct numerical simulations of 3D compressible MHD turbulence were performed in order to study the relation between waves modes and coherent structures and the consequent energization of test particles. 
Moreover, the question of which is the main mechanism of this particle energization is rigorously discussed. 
In particular, using the same initial conditions, we analyzed the non-linear and linear evolution of a turbulent state along with the case of randomized phases. 
Then, the behavior of the linear and non-linear simulations were compared through the study of time evolution of particle kinetic energy and preferential concentration. 
Also, spatio temporal spectra were used to identify the presence of wave modes and quantify the fraction of energy around the MHD modes in linear and non-linear simulations. 
Finally, the variation of the correlation time of the external forcing is studied in detail along with the effect on the particle energization (and clustering) and the presence of wave modes. 
More specifically, particle energization tends to decrease when the fraction of linear energy increase, supporting the idea that energization by structures is the dominant mechanism for particle energization instead of resonating with wave modes as suggested by Fermi energization theory.    
\end{abstract}

\section{Introduction} \label{sec:intro}

One of the first and most influential explanations for the origin of cosmic ray radiation was proposed by Enrico Fermi in 1949 \citep{Fermi1949}.
In that work, he argued about particles that could interact and resonate with passing Alfvén waves to achieve high kinetic energy. 
This explanation was later refined into the Quasi Linear Theory (QLT), which gave clear conditions for this resonance \citep{Stix1992}.
In terms of particle energization, QLT provides an estimation for diffusion coefficients in momentum space.
However, these calculations rely on the assumption of weak turbulence \citep{chandran2008,chandran2005}, where each field is decomposed as a mean background value plus a small amplitude fluctuation described as a collection of weakly interacting waves.

For plasmas in a fully developed turbulent regime, as those present in astrophysical plasmas, these assumptions are not fulfilled and QLT is no longer adequate. Moreover, in the so-called strong turbulence regime, the main role in particle energization is played by self consistent structures \citep{Dmitruk2004, dmitruk2006a, Gonzalez_2017, Lemoine2021, pugliese2022, Pezzi2022, Balzarini2022}. Particles are able to exploit the electric fields present in these structures to obtain a net energization during a whole gyroperiod \citep{Dmitruk2004, pugliese2022}. In the presence of a guide field, this energization is mainly perpendicular for protons and other heavy ions. Recent theoretical \citep{Lemoine2021} and numerical studies \citep{Greco2014, Pezzi2022, pugliese2022} have shown that some structures are also able to capture protons, which greatly enhances their energization capabilities.

At larger spatial scales, the magnetohydrodynamic (MHD) model provides a comprehensive explanation of interplanetary space plasmas, including the solar wind and planetary magnetospheres \citep{G2010}. Within this framework, two distinct descriptions of the turbulence phenomenon can be considered. On one hand, the wave behavior, which describes field fluctuations as a collection of waves allowed by the MHD approximation, such as Alfvén waves or magneto-acoustic waves in the plain compressible case \citep[see,][]{G2006}. On the other hand, the well-known strong turbulence regime, in which a broadband spectrum of fluctuations with coherent structures and the presence of intermittency, including non-propagating fluctuations, is present in wavenumber and frequency spaces \citep[see,][]{P2022}. To analyze the wave versus strong turbulence regimes, a spatiotemporal spectrum analysis of the fields can be performed \citep[see,][]{Me2015,Cl2015}. The scenario emerging from this analysis is one in which both the wave and turbulent behaviors of the fluctuations could coexist without contradictions \citep{Dmi2009}. This is supported by in situ observations data, such as those of the solar wind, which show both wave and turbulent behaviors in fluctuations \citep{Barnes1979,TuMarsch1995,Sa2022}. It is worth mentioning that the spatiotemporal analysis have been used to investigate incompressible hydrodynamic and incompressible and compressible MHD turbulence from experimental and numerical data \citep{C2014,A2017a,L2016,L2019,Brodiano_2021,K2022}.

Charged test particles provide a useful representation of a small fraction of charged particles in a plasma that can be energized by the fields; however, neglecting the feedback of the particles into the electromagnetic field - which would require a kinetic description of the plasma - limits their application. Despite this, charged test particles are commonly used to study the energization of charged particles in different scenarios.
In any case, electromagnetic fields need to be prescribed and it is usual to do so in Fourier space \citep{Dalena2012, Tautz2013}.
Multiple models are available to provide the amplitudes simulating turbulent spectra such as Slab or 2D \citep{Shalchi2009}, while phases are usually taken as random and uncorrelated.
For example, in solar wind a combination of 20\% Slab / 80\% 2D is realistic at 1AU heliocentric distance \citep{Bieber1996}.
Another approach is to use electromagnetic fields obtained from the evolution of MHD equations \citep{Dmitruk2004, Dalena2014, Gonzalez2016, pugliese2022}.

One important question is whether the wave or strong turbulent picture of the background plasma is more pertinent for the energization of the charged particles. In the present work, we concentrate specifically on this issue, analyzing the behavior of charged test particles with different techniques to determine the effect of the full MHD evolution of the field versus an evolution where non-linearities are artificially suppressed, resulting in only linear (wave) behavior. Our study provides important insights into the behavior of charged particles in plasmas and how they are energized by the surrounding fields.

The paper is organized as follows: in section \ref{sec:theo}, we introduce the theoretical CMHD set of equations and we present the linearized and dimensionless version. Then, we show the dispersion relation of the MHD waves modes, within a briefly explanation of the spatio-temporal spectrum technique. Finally, we present the equations regarding charged test particles and describe the Voronoi Tessellation method, used to quantify the concentration of particles in space. In section \ref{sec:num_set}, we describe the numerical set up for non-linear and linear simulations along with some details about phase randomization for the linear runs and particle integration. In section \ref{sec:res}, we present our results. Finally, in \ref{sec:disc} we summarize our main findings.

\section{Theory}\label{sec:theo}
\subsection{Compressible MHD equations}\label{ssec:CMHDeq}

The three dimensional (3D) compressible MHD (CMHD) model is given by the mass continuity equation, the induction equation for the magnetic field, the momentum Navier-Stokes equation and a polytropic state equation (relating pressure and densisty) and involves fluctuations of the velocity field  $\bf{u}$, density $\rho$ and magnetic field $\bf{B}=\bf{B}_0+\bf{b}$, where  $\bf{B}_0$ is the mean field and $\bf{b}$ is
the fluctuating part, 
\begin{equation}
    \frac{\partial \rho}{\partial t}+\boldsymbol\nabla\cdot(\rho \mathbf{u})=0,
    \label{continuidad}
\end{equation}
\begin{equation}
    \frac{\partial \bf{B}}{\partial t}=\boldsymbol\nabla\times(\bf{u}\times \bf{B})+\eta \nabla^2 \bf{B},
    \label{induccion}
\end{equation}
\begin{equation}
     \frac{\partial \mathbf{u}}{\partial t}+\mathbf{u} \cdot{\boldsymbol\nabla} {\bf u} =-\frac{{\boldsymbol\nabla} P}{\rho}+\frac{\mathbf{J}\times\mathbf{B}}{4\pi \rho}+\frac{\mu}{\rho}\bigg[\nabla^ 2 \mathbf{u}+\frac{{\boldsymbol\nabla} ({\boldsymbol\nabla}\cdot{\bf u})}{3}\bigg],
     \label{NS}
\end{equation}
\begin{equation}
    \frac{P}{\rho^\gamma}=\text{constant}. 
    \label{Politrópica}
\end{equation}
Note that changes of the magnetic field absolute value $B_0$ involves modifying the relative amplitude between the initial fluctuations $b$ and $B_0$ (with the initial conditions untouched). In addition, $P$ is the scalar isotropic pressure, $\bf{J}=(4\pi/c)\boldsymbol\nabla\times\bf{B}$ is the electric current, $\gamma=5/3$ is the polytropic index, and $\mu$ and $\eta$ are the dynamic viscosity and magnetic diffusivity, respectively. 
The main purpose of these last terms is to dissipate energy at scales smaller than MHD scales, while allowing us to study with an adequate scale separation compressible effects at the largest scales. 

The set of equations \eqref{continuidad}-\eqref{Politrópica} can be written in a dimensionless form in terms of a characteristic length scale $L_0$, a mean scalar density $\rho_0$ and pressure $P_0$, and a typical magnetic and velocity field magnitude $b_{rms}$ and $v_0=b_{rms}/\sqrt{4\pi\rho_0}$ (i.e., the r.m.s.~Alfv\'en velocity), respectively. Then, the unit time is $t_0=L_0/u_{rms}$, which for MHD becomes the Alfv\'en crossing time. The resulting dimensionless equations are,
\begin{equation}
    \frac{\partial \rho}{\partial t}+\boldsymbol\nabla\cdot(\rho \mathbf{u})=0,
    \label{continuidad_2}
\end{equation}
\begin{equation}
     \frac{\partial {\mathbf{u}}}{\partial t}+{\mathbf{u}} \cdot{\boldsymbol\nabla} {\bf u} =-\frac{1}{\gamma M_s^2}\frac{{\boldsymbol\nabla} P}{\rho}+\frac{{\mathbf{J}}\times{\mathbf{B}}}{\rho}+\frac{1}{\rho R_e}\bigg[\nabla^ 2 {\mathbf{u}}+\frac{{\boldsymbol\nabla} ({\boldsymbol\nabla}\cdot{\bf u})}{3}\bigg],
     \label{NS_2}
\end{equation}
\begin{equation}
    \frac{\partial {\bf B}}{\partial t}=\boldsymbol\nabla\times({\bf u}\times {\bf B})+\frac{1}{R_m} \nabla^2 {\bf B},
    \label{induccion_2}
\end{equation}
\begin{equation}
    P = \rho^\gamma
    \label{Politrópica_2}
\end{equation}
where $M_s=v_0/C_s$ is the sonic Mach number and $C_s^2=\gamma P_0/\rho_0$ is the sound speed.
The kinetic and magnetic nominal Reynolds numbers are also defined as $R_e = L_0v_0/\nu$ and $R_m=L_0v_0/\eta$, respectively, with $\nu=\mu/\rho_0$ the kinematic viscosity.
Considering a static equilibrium ($u_0=0$) with homogeneous external magnetic field ${\bf{B}_0}=B_0\vectorunit{z}$, a constant density $\rho_0$, and a constant pressure $P_0$, we can linearize Eqs.~\eqref{continuidad_2}-\eqref{Politrópica_2},

\begin{equation}
    \frac{\partial \rho}{\partial t}+\boldsymbol\nabla\cdot \mathbf{u}=0,
    \label{continuidad_lin_2}
\end{equation}
\begin{equation}
     \frac{\partial \mathbf{u}}{\partial t} =-\frac{1}{M_s^2}{\boldsymbol\nabla} \rho+\mathbf{J}\times\mathbf{B}_0+\frac{1}{\rho_0R_e}\bigg[\nabla^ 2 \mathbf{u}+\frac{{\boldsymbol\nabla} ({\boldsymbol\nabla}\cdot{\bf u})}{3}\bigg],
     \label{NS_lin_2}
\end{equation}
\begin{equation}
    \frac{\partial \bf{b}}{\partial t}=\boldsymbol\nabla\times(\mathbf{u}\times \mathbf{B}_0)+\frac{1}{R_m} \nabla^2 \bf{b},
    \label{induccio_lin_2}
\end{equation}
where the polytropic equation \eqref{Politrópica} was used to replace the pressure in the equations.

\subsection{Compressible MHD waves modes}\label{ssec:CMHDwaves}

In order to study the CMHD normal modes, we used Eqs. \eqref{continuidad_lin_2}-\eqref{induccio_lin_2} and obtain the dispersion relation $\omega(\bf{k})$ of small amplitude waves propagation in a plasma. It is straightforward to show that there are
three independent propagating modes (or waves), which correspond to the so-called Alfv\'en waves (A), fast (F), and slow (S) magnetosonic waves \citep[e.g.,][]{fitzpatrick2014plasma},
\begin{align}\label{Alven_wave} 
    \omega^2_A(k)&=k_{\parallel}^2u_A^2, \\ \label{magnetosonic}
    \omega_{F,S}^2(k)&=k^2u_A^2\left[\frac{(1+\beta)}{2}\pm \sqrt{\frac{{(1+\beta)}^2}{4}-\beta\left(\frac{k_{\parallel}}{k} \right)^2}\right],
\end{align}
where $\beta={(C_s/u_A)}^2$ is the plasma beta, i.e., the ratio of plasma pressure to magnetic pressure, which can expressed as $\beta=1/(M_sB_0)^2$) with $u_A=B_0/\sqrt{4\pi\rho_0}$ the Alfv\'en velocity, $C_s$ as defined above and $k=|{\bf k}|=\sqrt{{\bf k}_{\parallel}^2+{\bf k}_{\perp}^2}$ with ${\parallel}$ and $\perp$ the wavenumber component along and perpendicular to the external magnetic field, respectively. It is worth noting that $B_0$ is a parameter and has no relation with the initial fluctuations of the system. On one hand, Alfv\'en waves are incompressible fluctuations transverse to the external magnetic guide field ${\bf B}_0$. On the other hand, both fast and slow modes, unlike Alfv\'en modes, carry density fluctuations and their magnetic field perturbations have longitudinal and transverse components. In the case of fast modes, the magnetic field and the plasma are compressed by the wave motion, such that the restoring force is large and hence the frequency and the propagation speed are high. While, for slow modes, the magnetic field oscillation and the pressure variation are anti-correlated with each other such that the restoring force acting on the medium is weaker than that for the fast mode. For this reason the frequency and the propagation speed are the lowest among the three MHD waves branches.  
Note that for the perpendicular propagation (i.e., $k_{\parallel}=0$ and $k_{\perp}\ne 0$), the Alfv\'en and slow modes become non-propagating modes (i.e., $\omega_{A,S}=0$) and are degenerate, but they can be distinguished using their different polarization, since $\delta B_{\parallel A}=0$ and $\delta B_{\parallel S}\ne 0$ \citep{A2017a}. Finally, it is worth mentioning that, we adopt the assumption that energy concentrated closely to the linear dispersion relation can be explained by linear and weak turbulence theories, while any spread away from those linear curves is a sign of strong turbulence that requires fully nonlinear theories to be understood \citep[see,][]{A2017a,Brodiano_2021}. 

\subsection{Spatio-temporal spectrum}\label{ssec:sptempspec}

The spatio-temporal spectrum consists of calculating the complete spectrum in wavenumber and frequency for all available Fourier modes in a numerical simulation or an experiment \citep{C2014,Cl2015}. As a result, it can distinguish between modes that satisfy a given dispersion relation (and are thus associated with waves) from those associated with nonlinear structures or turbulent eddies, and quantify the amount of energy carried by each of them at different spatial and temporal scales. It is worth mentioning, that this method does not require the pre-existence of wave modes or eddies. In the following, the spatio-temporal magnetic energy spectral density tensor is defined as:
\begin{equation}\label{ten_spec}
    E_{ij}({\bf k},\omega)=\frac{1}{2}\vectorunit{{\bf B}}_i^*({\bf k},\omega)\vectorunit{{\bf B}}_j({\bf k},\omega),
\end{equation}
where $\vectorunit{\bf B}_i(\textbf{k},\omega)$ is the Fourier transform in space and time of the $i$-component of the magnetic field ${\bf B}({\bf x},t)$ and the asterisk implies the complex conjugate. The magnetic energy is associated with the trace of $E_{ij}(\textbf{k},\omega)$.

As the external magnetic field $\textbf{B}_0$ in the simulations
points in $\vectorunit{z}$, in practice, we will consider either $i=j=y$ or $i=j=z$, to identify different waves based on their polarization (either transverse or longitudinal with respect to the guide field). In all cases, the acquisition frequency was at least two times larger than the frequency of the fastest wave, and the total time of acquisition was larger than the period of the slowest wave in the system. 
It is worth mentioning that spatio-temporal spectra have been used before in numerical simulations and experiments of rotating turbulence \citep{C2014}, stratified turbulence \citep{di2015}, quantum turbulence \citep{di2015b}, and IMHD turbulence simulations \citep{Me2015,Me2016,L2016}, compressible MHD turbulence \citep{A2017a,Brodiano_2021} and in spacecraft observations \citep{Sahraoui2003,Sahraoui2010}. Quantifying the relative presence of waves and/or nonlinear structures is the main outcome expected from the spatio-temporal analysis. In particular, we use the spatio-temporal spectra to search for the presence of waves and quantify the energy found near their respective dispersion relation in a linear and nonlinear MHD turbulence. 

\subsection{Test particle equations}\label{ssec:particles}

We studied charged particles evolving with the dynamical MHD fields, but those fields are unaffected by the particles (i.e., test particles). The dynamics of a test particle in an electromagnetic field are given by the non-relativistic equation of motion:
\begin{equation}{\label{eq:newton}}
\frac{d\mathbf{r}}{dt} = \mathbf{v}, \quad \frac{d\mathbf{v}}{dt} = \alpha \left( \mathbf{E} + \mathbf{v}\times\mathbf{B} \right),
\end{equation}
where the electric field $\mathbf{E}$ is obtained from Ohm's Law and its dimensionless (using a characteristic electric field $E_0=v_0B_0/c$) expression is,
\begin{equation}{\label{eq:ohm}}
\mathbf{E} = \frac{\mathbf{J}}{R_m} - \mathbf{u}\times\mathbf{B}.
\end{equation}

While a more general version of this law includes the electronic pressure (i.e., $\nabla P_e/\rho$) and the Hall term (i.e, $\mathbf{J}\times\mathbf{B}/\rho$), here we neglect them in order to simplify the analysis and interpretation of the results and
to maintain consistency with the compressible MHD equations introduced in the previous sections. Several works have studied the effect of these terms on test particles \citep{dmitruk2006a, pugliese2022, Balzarini2022}.
As such, density fluctuations will not directly affect particle motion through Eq. \eqref{eq:ohm}, but could allow the existence of magnetosonic wave modes for the particles to interact with.

The parameter $\alpha$ is related to the charge-to-mass ratio and represents the gyrofrequency $\omega_g$ in a magnetic field of intensity $b_{rms}$ \citep{pugliese2022}. Its inverse value $1/\alpha$ represents the nominal gyroradius $r_g$ (in units of $L_0$) for particles with velocity $v_0$ in a magnetic field $b_{rms}$, and gives a measurement of the range of scales involved in the system (from the outer scale of turbulence to the particle gyroradius). For plasmas with a strong magnetic field amplitude $B_0$, we can define an average gyroperiod $\tau_p = 2\pi/\omega_g = 2\pi/\alpha B_0$. In the case of protons, we can relate $\alpha$ to the MHD field parameter through,
\begin{equation}{\label{eq:def_alpha}}
\alpha = \frac{L_0}{d_{i}},
\end{equation}
where $d_{i} = m_p c/\sqrt{4\pi\rho_0e^2}$ is the proton inertial length, with $m$ and $e$ the proton mass and charge, respectively. Furthermore, in the present paper, we identify the proton inertial length with the dissipation scale $d_{i}=l_d$, given the solar wind observations supporting $d_{i}\sim l_d$ \citep[e.g.,][]{Leamon1998}.

\subsection{Particle energization and clustering}\label{ssec:clustering}

Recently, \citet{pugliese2022} reported a link between high energization of test particles and a preferential concentration (clustering) for test particles.
In the present work, to quantify this preferential concentration, we employ the Voronoi tessellation method and compare volume statistics against a Random Poisson Process (RPP) \citep[see,][]{Monchaux2010, Obligado2014, Reartes2021}. In particular, this process provides a cell for each particle, whose volume $\mathcal{V}_i$ can be interpreted as the inverse of the local particle density. Therefore, by studying the statistics of these volumes, we can quantify any preferential concentration that may be present in the system. In addition, we compared against those of a uniform distribution in space, known as RPP \citep{UHLMANN2020}.
One of these statistic tools is the standard deviation of the volumes $\sigma_{\mathcal{V}}$, which should coincide with $\sigma_{RPP}$ for a uniform distribution and increases as the preferential concentration increases.

In the aforementioned work it was found that particles accumulated in regions where $\nabla_\perp\cdot\mathbf{u}_\perp = \partial_xu_x+\partial_yu_y < 0$.
Through Eq. \eqref{eq:ohm} this also implies $(\nabla\times\mathbf{E})_z<0$ and thus a clockwise rotating electric field.
Therefore, in this regions the electric force and particle gyration are aligned and there is positive energization after a whole gyroperiod.
This, in combination with the trapping effect of $\nabla_\perp\cdot\mathbf{u}_\perp < 0$, makes this mechanism very effective at energizing particles.

\section{Numerical set up}\label{sec:num_set}
\subsection{Nonlinear simulations}\label{ssec:nonlinsim}

The nonlinear simulations are performed by numerically solving Eqs. \eqref{continuidad}-\eqref{Politrópica} using a pseudospectral method with periodic boundary conditions in a cube of size $L_{box}=2\pi$.
This scheme ensures exact energy conservation for the continuous time spatially discrete equations \citep{Mininni2011}, where we use a spatial resolution of $N^3=512^3$ Fourier modes.
Time integration is achieved through a second order Runge-Kutta method. Aliasing is removed using the two-thirds truncation method \citep{Orszag1971}, such that the maximum wavenumber resolved is $\kappa \equiv N/3 = 170$.
To ensure the resolution of the smallest scales ($\kappa > k_d$), we use kinematic and magnetic Reynolds numbers of $R_e=R_m\approx2400$.
Here, $k_d = (\epsilon_d /\nu^3)^{1/4}$ is the Kolmogorov dissipation wavenumber (which defines the dissipation scale $l_d = 2\pi/k_d$), with $\nu=\mu/\rho_0$ the kinetic viscosity and $\epsilon_d$ the energy dissipation rate.

In order to reach a steady turbulent state, the system is forced using mechanical and electromotive forces $\mathbf{f}$ and $\nabla\times\mathbf{m}$, respectively. 
These forces are generated with random phases in the Fourier $k$-shells $2\leq |\mathbf{k}|\leq 3$ every correlation time $\tau_c$, which is also a controlled parameter in the simulations.
Forces at intermediate times are obtained by linearly interpolating the previous and next steps. 
We repeat this procedure for three different values of $\tau_c$, yielding the three different stationary states summarized in Table \ref{tab:nl_values}.
The length and time scales are defined individually for each simulation.
For the characteristic length scale $L_0$, we use the energy containing scale $L_0 = 2\pi \int (E(k)/k) dk / \int E(k) dk$ where $E(k)$ is the isotropic energy spectrum.
For the velocity scale, we use the Alfvén velocity of the magnetic field fluctuations $v_0=b_{rms}/\sqrt{4\pi\rho_0}$. 
For the time scales, we alternatively use the large eddy turnover time $t_0=L_0/v_0$ and the particle gyroperiod $\tau_p = 2\pi/\alpha B_0$, depending on whether we analyze field or particle properties.
Finally, as seen in Table \ref{tab:nl_values}, we have a ratio of mean magnetic field to magnetic field fluctuation ($B_0/b_{rms}$) of $1:9$, similar to values reported in the solar wind \citep{H2017,A2022}.

\begin{deluxetable*}{cccccc}
\tablecaption{Global quantities for nonlinear simulations}
\tablewidth{0pt}
\tablehead{
\colhead{Run} & \colhead{$\tau_c/\tau_p$} & \colhead{$L_{box}/L_0$} & \colhead{$L_0/l_d$} & \colhead{$B_0/b_{rms}$} & \colhead{$u_{rms}/v_0$}
}
\startdata
NL1 & $1.146\times10^1$    & $2.53$ & $59.79$ & $8.97$ & $1.73$ \\
NL2 & $1.146\times10^0$    & $2.43$ & $54.69$ & $9.16$ & $1.71$ \\
NL3 & $2.865\times10^{-1}$ & $2.83$ & $65.86$ & $8.49$ & $1.49$ \\
\enddata
\tablecomments{Energy injection scale, dissipation scale, mean magnetic field and characteristic velocity obtained with different forcing correlation times. All magnitudes are similar, allowing us to compare the simulations on equal footing.}
\label{tab:nl_values}
\end{deluxetable*}

\subsection{Linear simulations}\label{ssec:linsim}

The linear simulations are performed in the same way as the nonlinear ones but we cancel all the non-linear terms in 
the MHD equations, as shown in Eqs.~\eqref{continuidad_lin_2}-\eqref{induccio_lin_2}.
In the absence of nonlinear terms, there is no energy cascade from the injection scale $L_0$ to smaller scales.
Therefore, modes with wavenumber $\mathbf{k}$ outside the forced shell would quickly vanish in the presence of the dissipation terms in Eqs.~\eqref{NS_lin_2}-\eqref{induccio_lin_2}.
To counter this, we evolve the fields in Eqs.~\eqref{continuidad_lin_2}-\eqref{induccio_lin_2} without forcing and without dissipation (i.e. $R_e,R_m\to\infty$), thus ensuring a constant energy spectra.
Furthermore, we use a fourth order Runge-Kutta scheme for temporal integration, as the second order scheme becomes unstable in the absence of dissipation terms.

\begin{figure}[h]
    \centering
    \includegraphics[width=0.85\columnwidth]{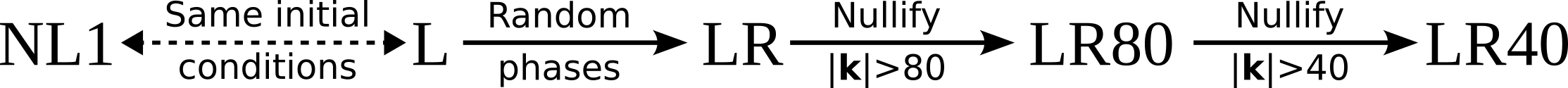}
    \caption{Schematic diagram showing how the initial conditions of each linear run are constructed from the initial conditions of the NL1 run.}
    \label{fig:linsimsumm}
\end{figure}

The initial conditions for the linear simulations are given by different variations of the initial conditions used in NL1, as summarized in Fig. \ref{fig:linsimsumm}. 
For the L run, we use these initial conditions unchanged, while for the LR run we perform a phase randomization.
We achieve this by transforming each Fourier coefficient $\psi_\mathbf{k} \mapsto e^{i\phi_\mathbf{k}}\psi_\mathbf{k}$ for all the scalar fields $\rho$, $u_j$ and $b_j$, where $\phi_\mathbf{k}$ are random phases independently chosen for each $\mathbf{k}$ and $\psi$.
We then enforce the necessary conditions on the resulting fields (i.e. hermiticity, Coulomb gauge).
These first two runs are built to have the same energy spectra as NL1, but thoroughly destroy any cross-field correlation and structures present in the initial conditions of NL1 for the LR run \citep{Alexakis2007}.

The final two runs LR80 and LR40 use the same Fourier coefficients as LR, but imposing $\psi_{\mathbf{k}} = 0$ for $|\mathbf{k}|>80$ and $|\mathbf{k}|>40$, respectively.
As this process slightly reduces the total energy, we compensate by uniformly re-scaling all the Fourier coefficients, thus preserving any power-law behaviour.

\subsection{Particle integration}\label{ssec:partint}

Initially, particles were uniformly distributed in the box and with a Gaussian velocity distribution function, with a root mean square (rms) of $\langle v_i^2 \rangle^{1/2}\approx 1.2v_0$. As shown in Table \ref{tab:nl_values}, this value lies between the Alfvén velocity $v_0$ and the rms value of the velocity field $u_{rms}$.
Furthermore, we chose $\alpha=60$ for the particles, which is consistent with the values of $L_0/l_d$ for all simulations.
The numerical integration of Eq. (\ref{eq:newton}) was done by a Runge-Kutta method with the same order as that used for field integration (see above). 
The values of the fields at each particle position are obtained by cubic splines in space from the three-dimensional grid of the MHD simulation.
The particle trajectories were integrated for $\sim300\tau_p$.

\section{Results} \label{sec:res}

\subsection{Linear vs nonlinear}\label{ssec:linvsnonlin}

\begin{figure}
    \centering
    \includegraphics[width=\columnwidth]{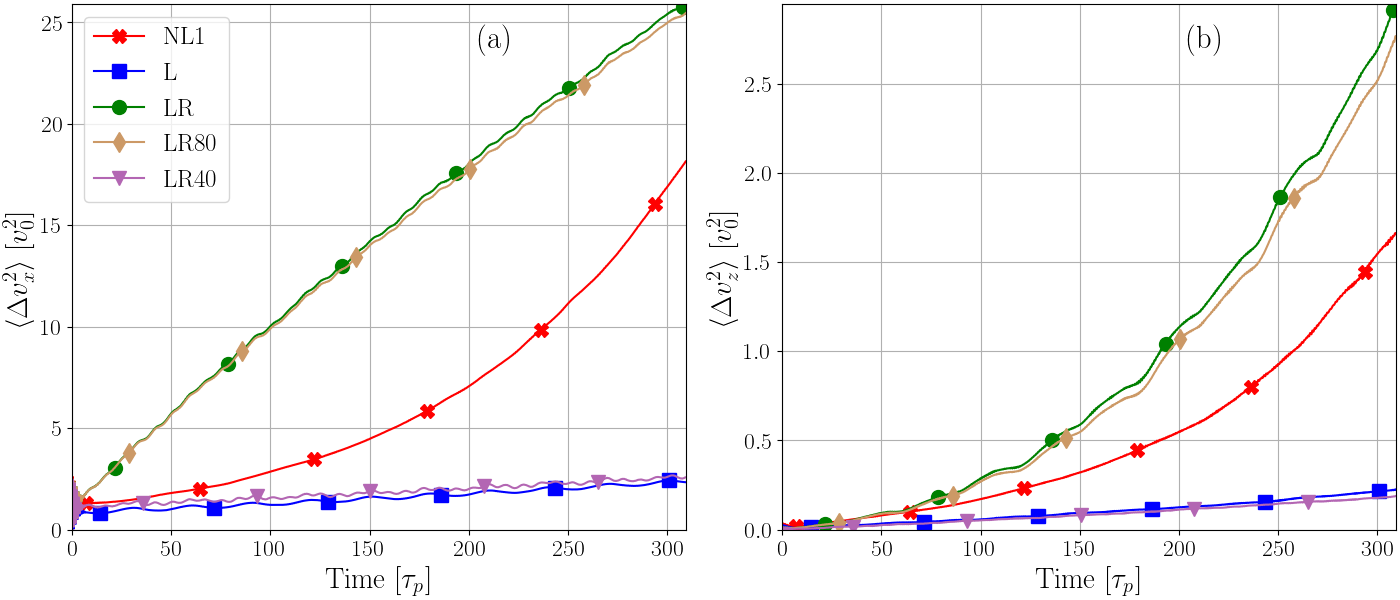}
    \caption{Time evolution of the mean kinetic energy for the perpendicular (left) and parallel (right) components, for the non-linear and linear simulations. For the time scale, we used the particle gyroperiod $\tau_p$.} \label{fig:energy_l_vs_nl}
\end{figure}

In this section, we compare the dynamics of test particles in the nonlinear run NL1 versus the dynamics in the linear counterparts L, LR, LR80 and LR40. 
We begin with the evolution of the mean particle energy, showing separately the parallel and perpendicular component to the magnetic guide field in Fig. \ref{fig:energy_l_vs_nl}(a) and Fig. \ref{fig:energy_l_vs_nl}(b), respectively.
As it is known for protons, energization is much higher in the perpendicular direction (represented here by the $x$ component) with respect to the parallel direction (represented here by the $z$ direction).
For both directions, it is clear that the linear case L has significantly smaller energization when compared to the full nonlinear simulation NL1.
However, for the phase randomizing (LR run), the energization increases greatly and surpasses that of NL1 run.
Truncating the spectra at $|\mathbf{k}|=80$ (LR80 run) has very little effect, but truncating at $|\mathbf{k}|=40$ (LR40 run) greatly reduces the energization, obtaining values similar to the linear L run.
Furthermore, in the perpendicular case shown in Fig. \ref{fig:energy_l_vs_nl}(a) all the linear runs seem to display a diffusive or subdiffusive behaviour (in momentum space) 
$\langle v_x^2 \rangle \sim t^\alpha$ with $\alpha \lesssim 1$.
This fact suggests that the underlying mechanism for energization (in the linear case) is analogous to that of Brownian motion, with delta correlated energy increments.
In this analogy, particles would have a very strong and time localized interaction with the fields. 
In the context of QLT, this very strong interaction could be understood as a resonance with some specific wave present in the system.
This high energization quickly removes the particle from the resonance condition within a few gyroperiods, thus ensuring the interaction to be localized in time.

\begin{figure}
    \centering
    \includegraphics[width=1\columnwidth]{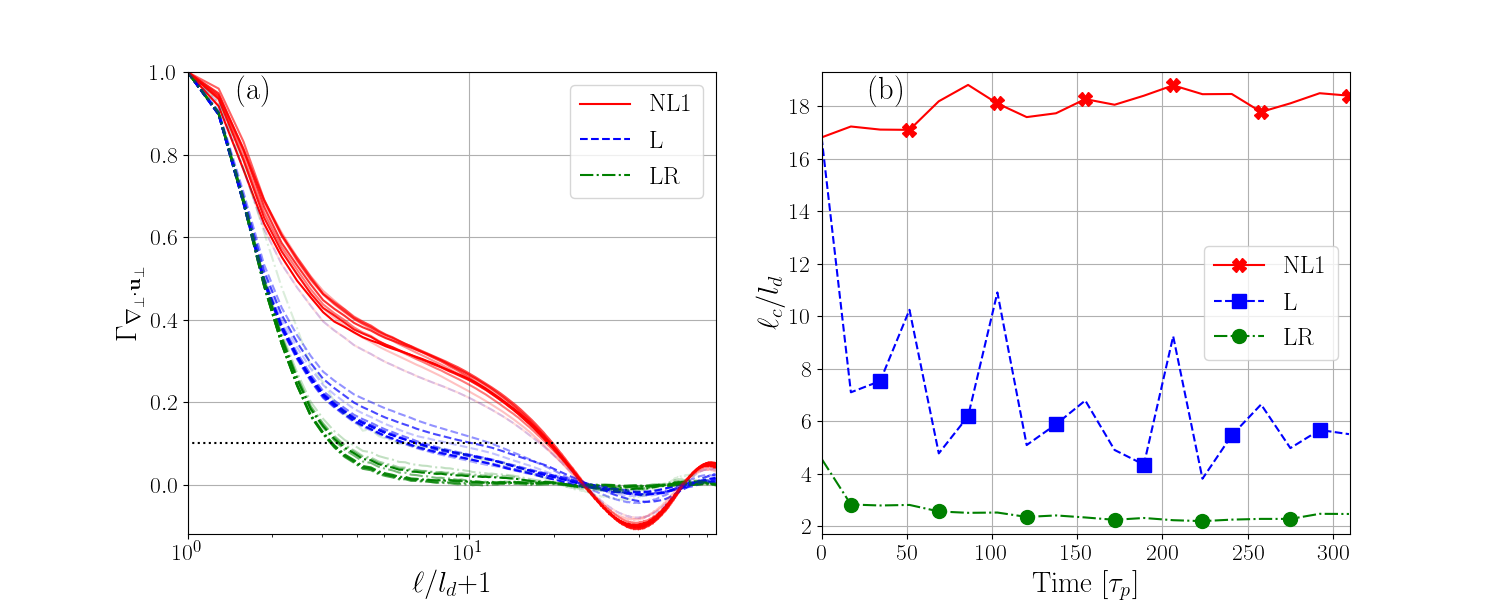}
    \caption{(a) Radial two-point autocorrelation $\Gamma$ of $\nabla_\perp\cdot\mathbf{u}_\perp$ in the plane perpendicular to the magnetic guide field at multiple times (later times are represented by darker colors) as a function of the displacement length $\ell$ normalized by the dissipation scale $\ell_d$, for non-linear and linear simulations. (b) Correlation length $\ell_c$, normalized by $\ell_d$, as a function of time; the correlation length $\ell_c$ is obtained as the displacement length $\ell$ at which the autocorrelation $\Gamma$ drops below $10\%$ of its maximum value (indicated by the dotted line in part (a) of this figure) . }
    \label{fig:corr_l_vs_nl}
\end{figure}

\begin{figure}
    \centering \includegraphics[width=0.5\columnwidth]{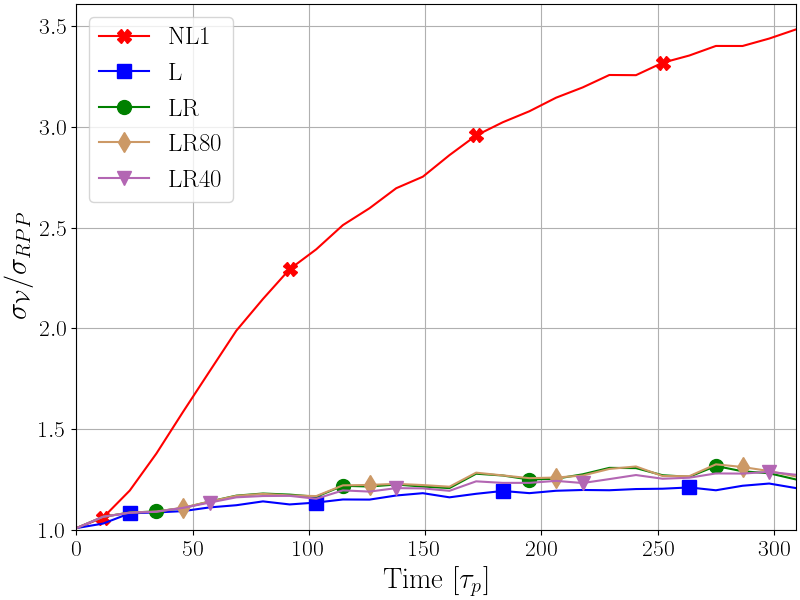}
    \caption{Standard deviation of the Voronoi volumes normalized by the standard deviation for a uniform distribution, as a function of time for non-linear and linear simulations.}
    \label{fig:sigmas_l_vs_nl}
\end{figure}

On the other hand, energization in the nonlinear case is superdiffusive, which is related to interaction between particles and self-consistent structures present in the plasma, as discussed in Section \ref{ssec:clustering}.
With this in mind, we could relate the drop in energization as NL1$\to$L to the disappearance of structures in the plasma due to the linear evolution.
In Eqs.~\eqref{Alven_wave} and \eqref{magnetosonic} we see that waves in the CMHD model are indeed dispersive and as such they will destroy any structure initially present in the system.
To investigate this, we calculated the radial two-point autocorrelation $\Gamma$ of $\nabla_\perp\cdot\mathbf{u}_\perp$ in the plane perpendicular to the magnetic guide field.
In Fig.~\ref{fig:corr_l_vs_nl}(a) we show the autocorrelation at multiple times (darker colors represent later times).
From these curves, we calculated the correlation length $\ell_c$, defined in this case as the displacement at which the autocorrelation drops below $10\%$ (horizontal dotted black line).
In Fig. \ref{fig:corr_l_vs_nl}(b) we show the autocorrelation length $\ell_c$ as a function of time, where we note that the NL1 run holds $\ell_c\approx18l_d$ during the whole simulation.
On the other hand, the linear run L starts with a similar value to the NL1 run but very quickly drops to $\ell_c\approx8l_d$ and then slowly decreases toward that of the LR run ($\ell_c\approx 2l_d$), which by construction should have negligible correlation (see Sec. \ref{ssec:linsim}).
Therefore, this confirms that the linear evolution eliminates most structures initially present in the field in less than $20\tau_p$, preventing particles from exploiting the energization mechanism described in Sec. \ref{ssec:clustering}.
To further check this claim, in Fig. \ref{fig:sigmas_l_vs_nl} we show the standard deviations of the Voronoi volumes as a function of time.
We see that the NL1 run quickly increases its clustering while all the linear runs remain very close to uniformity.
The slight increase in their $\sigma_\mathcal{V}$ could mean that this mechanism still survives in some small scale regions, but given the diffusive energization of Fig. \ref{fig:energy_l_vs_nl} we could consider this effect secondary at best.

We turn now to the comparison of the linear energization runs between themselves.
Having discarded structure interaction in the linear cases, we are left only with wave-particle resonant interaction.
The first observation is that the phase randomization L$\to$LR seems to vastly increase particle energization.
This fact mainly shows the importance of the phase mixing hypothesis in QLT.
The L run clearly does not fulfill this hypothesis, as its initial conditions are derived from a fully nonlinear turbulent simulation and as such display high phase correlation.
Alternatively, we could visualize the wave-particle interaction as a conjunction of resonance in terms of frequency and an initial alignment between the field and particle velocity.
This alignment is more easily achieved under the phase mixing hypothesis, while for a turbulent state phases do not actually fill every possible value at all frequencies.

\begin{figure}
    \centering
    \includegraphics[width=1\columnwidth]{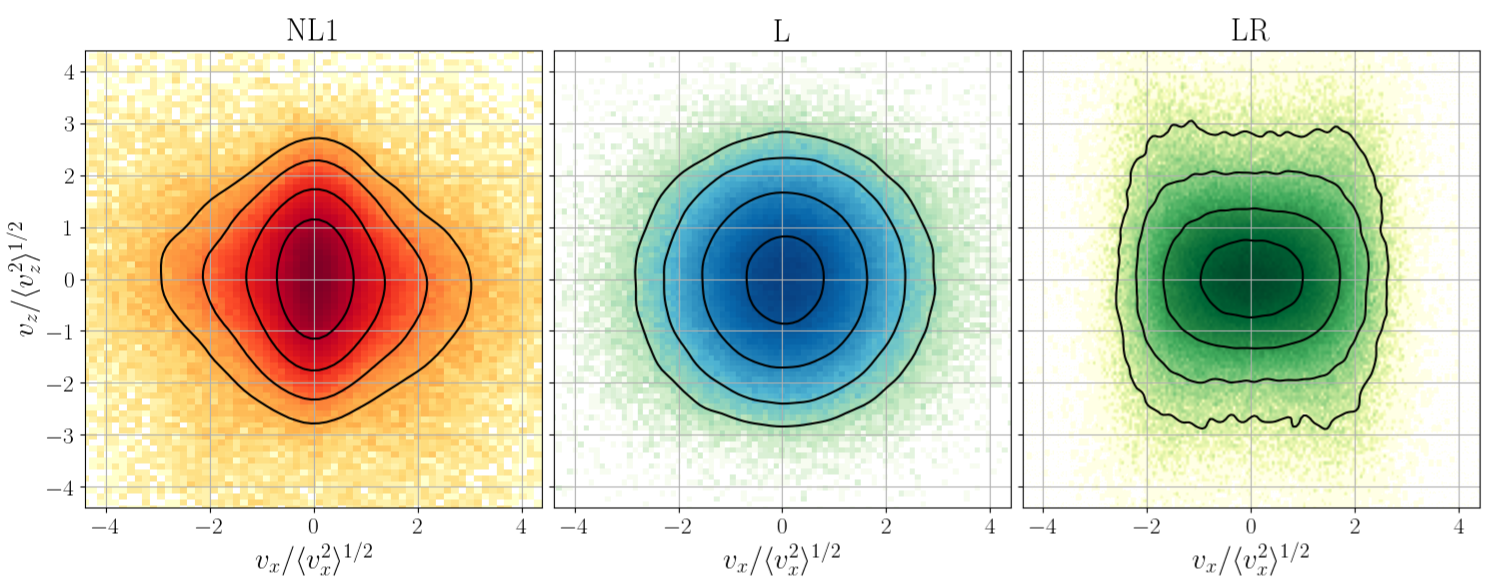}
    \caption{Two-dimensional PDF of particle velocities normalized by its standard deviation at the end of the simulation for the NL1, L and LR runs, respectively. The contours display different geometry, implying different underlying mechanism for energization.}
    \label{fig:2d_vel_hist}
\end{figure}

We can visualize these different energization mechanisms by computing the two-dimensional PDF of particle velocities at the end of each simulation.
The clear geometrical difference of each PDF shown in Fig. \ref{fig:2d_vel_hist} suggests that each mechanism is fundamentally different, as the data is normalized to the unit variance.
Close to the origin, all distributions are elliptic (although with different foci).
As we move from the origin, each distribution changes significantly. In particular, for the NL1 run, they tend to rhombi reminiscent of constant 1-norm (i.e., $|v_x|+|v_z|=$const.), suggesting that simultaneously high perpendicular and parallel energization is unlikely, as expected from this structure energization (high $v_z$ particles are difficult to trap).
For the L run, the shapes are clearly circular, corresponding to constant 2-norm ($v_x^2+v_z^2=$const.) or kinetic energy.
This is consistent with two independent Gaussian variables, as expected from a diffusive process in momentum space.
Finally, the LR run is less clear, with shapes similar to rectangles reminiscent of constant $\infty$-norm ($\max\{v_x, v_y\}=$const.).

By comparing the LR run with its truncated counterparts LR80 and LR40, we first see that the first truncation (i.e., LR$\to$LR80) has very little effect on particle energy, implying that particles resonate mainly with waves of $|\mathbf{k}|<80$.
Furthermore, the second truncation LR80$\to$LR40 reduces energization to even lower than the L run, showing that particles resonate mainly with waves of $|\mathbf{k}|>40$.
Two-dimensional velocity distributions for the LR80 and LR40 runs (not shown here) are very similar to those of LR and L in Fig. \ref{fig:2d_vel_hist}, respectively.
While the similarity between LR80 and LR is to be expected, that between LR40 and L is not obvious and seems to show that both runs share the same energization mechanism.
This last fact implies that changes in the phase distribution of the waves could be as important as changes in the energy spectrum.

\begin{figure}[]
    \centering
    \includegraphics[width=1.0\linewidth]{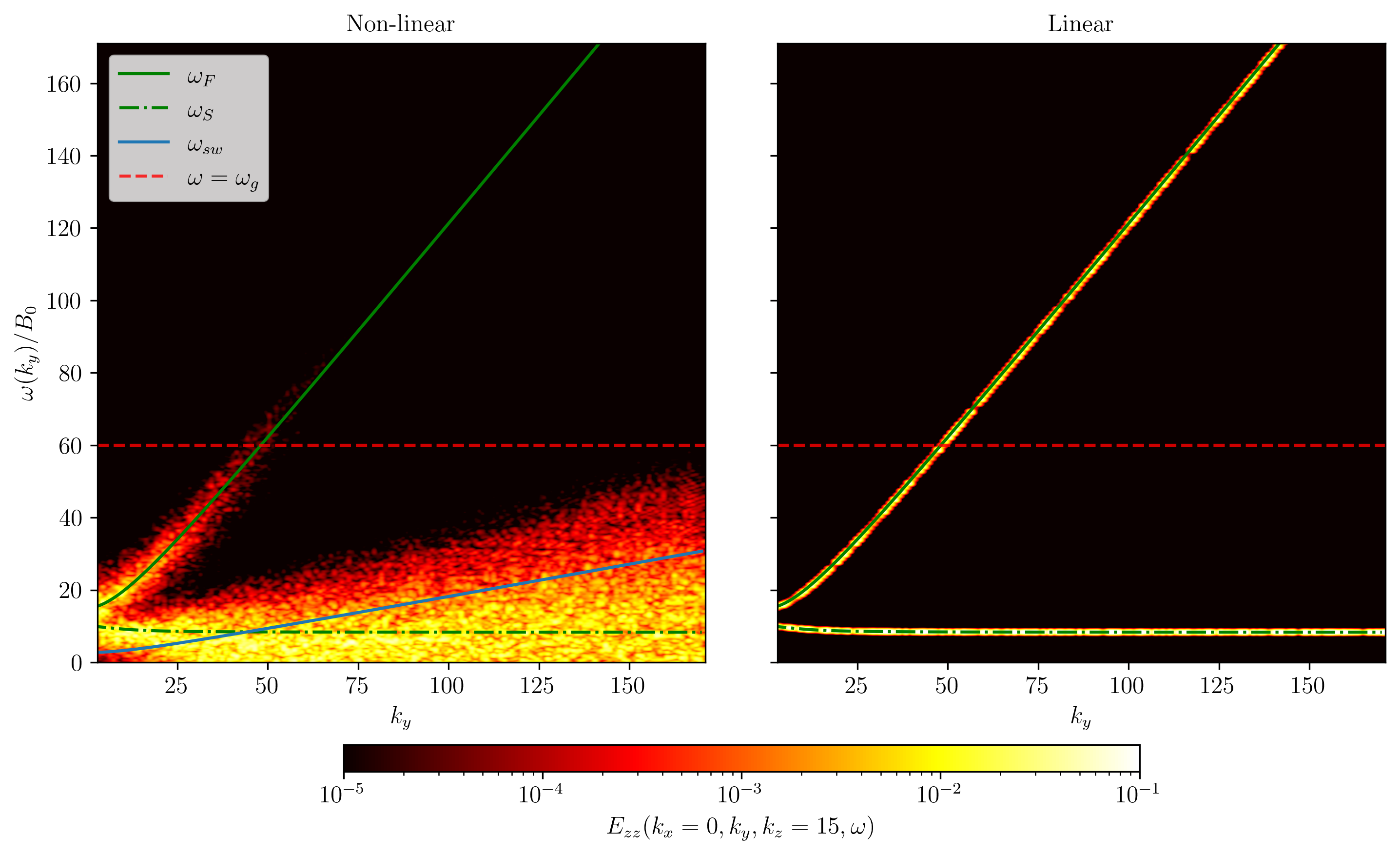}
    \caption{Spatio-temporal spectrum $E_{zz}(k_x=0,k_y,k_z=0)$ for the magnetic field fluctuations parallel to $B_0$, for the runs NL1 (left) and L (right). The spectrum is shown as a function of $\omega$ and $k_{y}$ for fixed $k_x=k_{\parallel}=0$. The green solid and the green dashed-dotted lines correspond to the linear dispersion relation of fast magnetosonic waves ($\omega_F$) and of slow magnetosonic waves ($\omega_S$), respectively. We include the particle gyrofrequency $\omega_g$ (dashed line) and the sweeping frequency $w_{sw}$ (blue line).} 
    \label{lin_vs_nolin}
\end{figure}

In order to distinguish between the presence of waves and structures in a plasma and to investigate which dominate the dynamics, the spatio-temporal spectra were used \citep[e.g.,][]{Cl2015,A2017a}. Therefore, we resolved the waves in time and space by choosing a very high frequency cadence to store the magnetic field, in particular we used $ dt = 2.5\times 10^{-4}$ as the temporal sampling rate. It is worth to mention that, we assumed that the energy concentrated around the linear dispersion relation can be explained by linear and weak turbulence theories \citep{chandran2005,chandran2008}, while any spread away from the dispersion relation is a sign of strong turbulence that requires fully nonlinear theories to be understood. Figure \ref{lin_vs_nolin} shows the spatio-temporal spectrum of the parallel magnetic field  $E_{zz}(k_x=0,k_y,k_z=0,\omega)$ for the NL1 and L runs (the rest of linear runs have similar spectra to the L run), i.e. we separate the linear from the nonlinear case. The dispersion relation for the fast and slow magnetosonic waves given by Eq. \eqref{magnetosonic} is shown in green solid and green dashed-dotted lines, respectively. Also, we include the particle gyrofrequency $\omega_g$ which lies between wavenumbers $40<|\mathbf{k}|<80$ and the characteristic sweeping frequency given by $\omega_{sw} \sim U_{rms}\sqrt{k_{\perp}^2+k_{\parallel}^2}$ \citep[see,][]{L2016}. 
As we expected, for the linear run L, the magnetic energy is located around the magnetosonic branches, while for the nonlinear run NL1, the energy accumulates mainly in the slow waves (with $\omega/B_0 = 0$ or non-propagating modes) for all wavenumbers. We also noted a small portion of energy spread along the fast branch for low wavenumbers. However, in the nonlinear case, most of the energy is spread across the spectrum due to its turbulent dynamic.  

Furthermore, the absence of sweeping in the linear case is also useful to understand the low energization in the L run.
As shown in Fig. \ref{fig:corr_l_vs_nl}, correlation quickly drops during linear evolution but not quite enough to reach the LR case, which implies that some structures may survive longer.
The absence of sweeping produced by the linear evolution means these surviving structures are not advected by the flow.
However, test particles are advected, thus making trapping more difficult and preventing any surviving structure from effectively energizing particles.

\subsection{The effect of \texorpdfstring{ $\tau_c$ }{tauc} on particle energization}\label{ssec:tcorr_part}

For this section, we repeated the analysis from the last section with the simulations NL2 and NL3, which differ from the NL1 run in the correlation time of the forcing $\tau_c$ (see Table \ref{tab:nl_values}).
Energization is qualitatively similar to that of NL1 (as shown in Fig. \ref{fig:energy_l_vs_nl}), but not quantitatively.
The same occurs for the correlation length and volume deviation $\sigma_\mathcal{V}$, as summarized in Table \ref{tab:nl_res} (energization and $\sigma_\mathcal{V}$ are calculated at the end of the simulation).
We see that both energization and correlation length decrease along with $\tau_c$, while $\sigma_\mathcal{V}$ displays no clear tendency and remains mostly similar.
First, the reduction of $~\ell_c/l_d~$ could be expected, as faster forcing (lower correlation time) prevent structures from being stable, either for lack of size or intensity.
Secondly, we could conclude that particles tend to accumulate similarly in all cases, but their ability to exploit energization mechanism diminishes for lower $\tau_c$.

Therefore, we need another way to quantify particle interaction with structures, including some dynamical information.
For this purpose, we performed Voronoi tessellation for each time step and calculated whether the particle was clustered.
We define a particle as clustered when its cell volume $\mathcal{V}$ is lower than some threshold value $\mathcal{V}_{th}$, which is obtained by comparing the cell volume PDF with that of a RPP (see references in Section \ref{ssec:clustering} for more details).
As seen in Fig. \ref{fig:sigmas_l_vs_nl}, clustering takes some time to settle, and as such the label clustered may not mean much initially.
Towards the end of the simulation, clustered is almost equivalent to trapped inside a structure. 
Then, we can determine when and where are particles accumulating and therefore calculate the amount of consecutive time they spent clustered.
We define a streak as an interval of (discrete) time for which the particle is clustered and its duration as $\mathcal{S}$.
In particular, we calculated the mean streak time $\langle\mathcal{S}\rangle$ for each simulation by averaging over all streaks from all particles and display it in the last column of Table \ref{tab:nl_res}.
Particles are clustered for around 2 full gyroperiods in all cases, but the mean time decreases with $\tau_c$.
This shows that while the clustering is very similar instantaneously for all runs, there is more exchange (between clustered and not clustered particles) in low $\tau_c$ simulations, suggesting clusters become more feeble and unstable with lower correlation time.
For lower values of $\tau_c$, the rapidly changing forcing may be experienced by particles as kicks that remove them from the structures trapping them.

\begin{figure}[]
    \centering
    \includegraphics[width=1.0\linewidth]{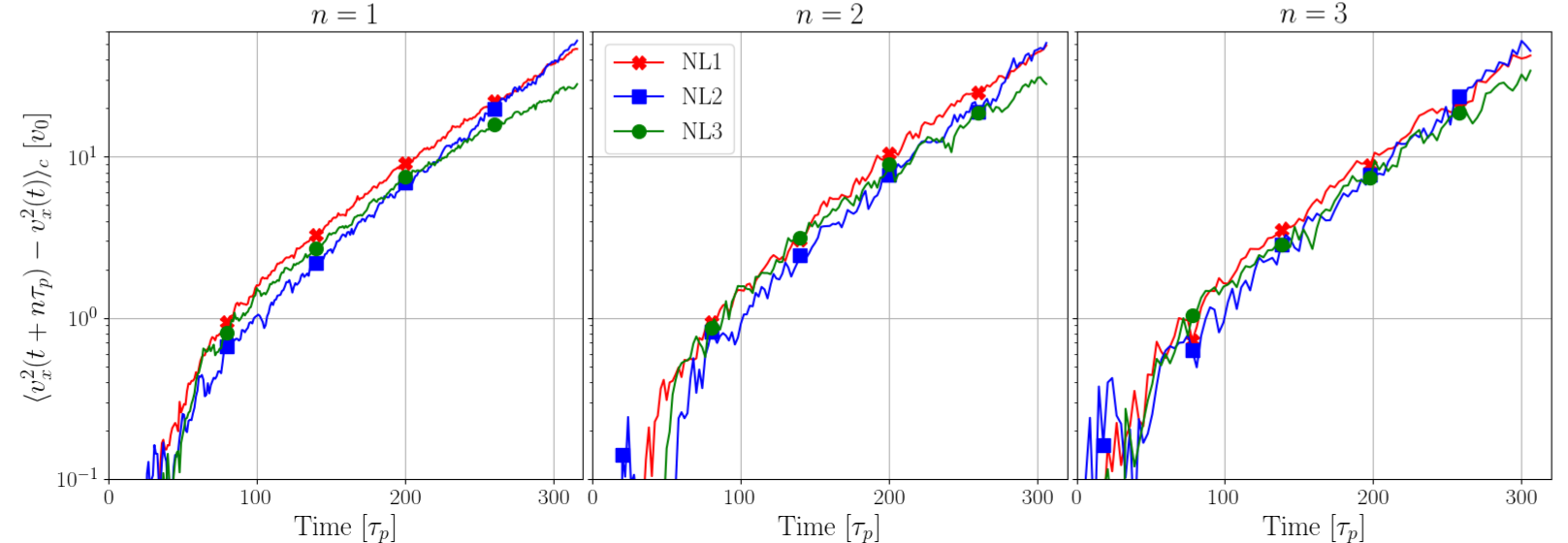}
    \caption{Mean of the structure function of the perpendicular particle energization for intervals with time duration $n\tau_p$ (where $n=1,2,3$). We selected particles that are clustered during this time.}
    \label{fig:deltaE_clus}
\end{figure}

To further perform this streak analysis, we related it directly with particle energization.
For this, we separated our data into intervals of duration time $n\tau_p$ ($n=1,2,3$) and we selected particles that are clustered during this time (whose streaks contained the interval under study).
For these particles, we calculated their mean perpendicular energization during these intervals and display it in Fig. \ref{fig:deltaE_clus}.
We can see that after $\sim 100\tau_p$ (long enough for clustering to settle), energization becomes exponential in time.

In order to understand this, we propose a very simple model for the interaction with these structures.
If we write the particle velocity in terms of its parallel and perpendicular components $\vb{v} = \vb{v}_\perp + \vb{v}_\parallel$, we can define the perpendicular energy as $\varepsilon_\perp = |\vb{v}_\perp|^2/2$.
Using Eq.~\eqref{eq:newton}, we could derive an evolution equation for the perpendicular energy $\varepsilon_\perp$,
\begin{equation}{\label{eq:energ_perp_teo}}
    \dv{\varepsilon_\perp}{t} = \alpha\left[ \vb{E}_\perp\cdot\vb{v}_\perp - \left(\vb{v}_\parallel \times\vb{v}_\perp\right)\cdot\vb{b}_\perp \right] \equiv \mathcal{P}_\perp - \mathcal{P}_{\cross}.
\end{equation}
The first term $\mathcal{P}_\perp=\alpha\vb{E}\cdot\vb{v}_\perp$ corresponds to net energization while the second relates to exchange with parallel energy, such as pitch angle scattering.
Averaging over one gyroperiod,
\begin{equation}
\expval{\mathcal{P}_\perp}_{\tau_p} = \frac{1}{\tau_p} \int_0^{\tau_p} \alpha\vb{E}_\perp\cdot\vb{v}_\perp \dd t \approx \frac{\alpha }{\tau_p}\oint_\mathcal{C} \vb{E}_\perp\cdot \dd \vb{l} = -\frac{\alpha }{\tau_p}\iint_\mathcal{S(C)} \curl\vb{E}_\perp\cdot \dd \vb{S}
\end{equation}
where we have approximated the trajectory on the perpendicular plane as circular and then used Stokes' Theorem, with a minus sign accounting for the clockwise circulation.
For a strong guide field, we could approximate Eq. \eqref{eq:ohm} as $\vb{E}_\perp \approx -\vb{u}_\perp\times\vb{B}_0$ and therefore the parallel component of its curl is $(\curl\vb{E}_\perp)_\parallel \approx B_0\boldsymbol{\nabla}_\perp\cdot\vb{u}_\perp$.
Expanding the integral at the lowest order in the gyroradius $R_g$ we achieve,
\begin{equation}
\expval{\mathcal{P}_\perp}_{\tau_p} \approx -\frac{\alpha^2 B_0^2}{2\pi}\iint_\mathcal{S(C)} \vb{\nabla}_\perp\cdot\vb{u}_\perp  \dd S \approx -\frac{\alpha^2 B_0^2}{2\pi} \pi R_g^2 \boldsymbol{\nabla}_\perp\cdot\vb{u}_\perp
\end{equation}
Using that $R_g = |\vb{v}_\perp|/\alpha B_0$ and recalling the definition of $\varepsilon_\perp$, we can substitute in Eq. \eqref{eq:energ_perp_teo} to obtain,
\begin{equation}
    \dv{\varepsilon_\perp}{t} \approx \lambda \varepsilon_\perp, \quad \lambda = -\boldsymbol{\nabla}_\perp\cdot\vb{u}_\perp,
\end{equation}
which for approximately constant $\boldsymbol{\nabla}_\perp\cdot\vb{u}_\perp<0$ predicts exponential increase for $\varepsilon_\perp$.

The slopes in Fig.~\ref{fig:deltaE_clus} represent $\lambda$ and are very similar, suggesting that clustered particles are energized at the same rate.
The difference in energization observed in Table \ref{tab:nl_res} must therefore be related to the time each particle spends clustered, as shown by $\langle\mathcal{S}\rangle$.
Particles can leave structures in basically 3 ways: (a) escaping vertically due to high $v_z$, (b) reaching the maximum allowed gyroradius, or (c) being pushed out by some fluctuation.
We disregard option (a) as particles in all simulations have very similar parallel energization (not shown).
Option (b) implies that the gyroradius $R_g=v_\perp/\alpha B_0$ is comparable to the size of the structures, which could be taken as $\ell_c$ and would be reasonable considering how it depends on $\tau_c$ (see Table \ref{tab:nl_res}).
We can calculate the required kinetic energy as $v_\perp^2 \sim (\alpha B_0 \ell_c)^2 \sim 1600 v_0^2$, which is achieved by practically no particle (less than $1$ particle in $10^4$) and as such cannot be the dominant cause.
This leaves option (c) as the last and main possibility, suggesting that structures are less robust against fluctuations and thus weaker.
In the wave/turbulence dichotomy, we could attribute this weakness to the prevalence of waves over structures in the system, to which we will dedicate the next section.

\begin{deluxetable*}{cccccc}
\tablecaption{Particle related quantities in nonlinear simulations}
\tablewidth{0pt}
\tablehead{
\colhead{Run} & \colhead{$\tau_c/\tau_p$} & \colhead{$\ell_c/l_d$} & \colhead{$\langle\Delta v_x^2\rangle/v_0^2$} & \colhead{$\sigma_\mathcal{V}/\sigma_{RPP}$} & $\langle\mathcal{S}\rangle/\tau_p$}
\startdata
$\text{NL1}$ & $1.146\times10^1$    & $18.0$ & $24.3$ & $1.87$ & $2.23$ \\
$\text{NL2}$ & $1.146\times10^0$    & $15.2$ & $20.3$ & $1.98$ & $2.13$ \\
$\text{NL3}$ & $2.865\times10^{-1}$ & $14.4$ & $15.6$ & $1.70$ & $1.85$ \\
\enddata
\tablecomments{Forcing correlation time, field correlation length, mean value of the perpendicular energization, volume deviation and clustered time of the particles for non-linear simulations. While correlation length, energization and clustered time decrease with $\tau_c$, volume deviation remains mostly constant.}
\label{tab:nl_res}
\end{deluxetable*}


\subsection{The effect of \texorpdfstring{ $\tau_c$ }{tauc} on spatio-temporal spectra}\label{ssec:waves}

\begin{figure}[]
    \centering
    \includegraphics[width=1.0\linewidth]{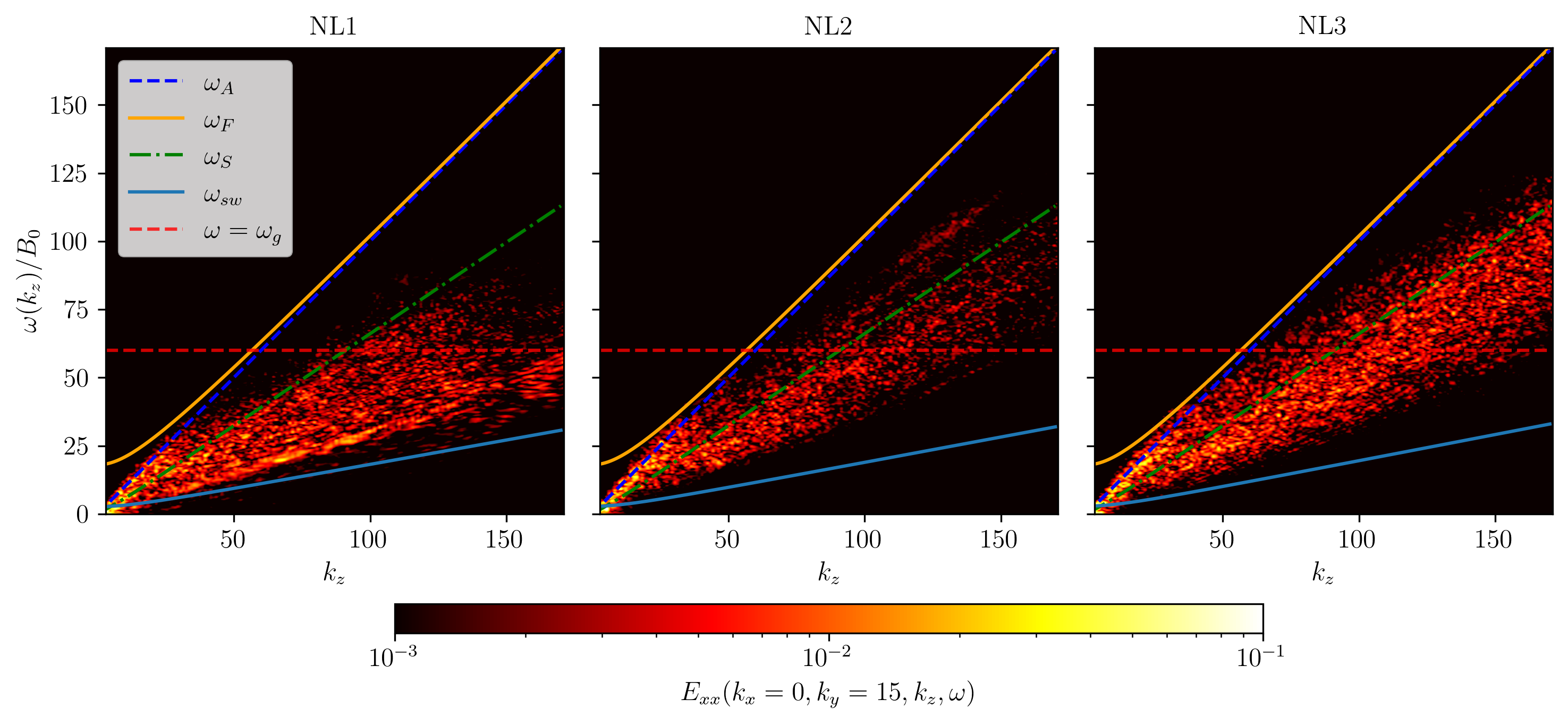}
    \caption{Spatio-temporal spectrum $E_{xx}(k_x=0,k_y=15,k_z)$ for the magnetic field fluctuations perpendicular to $B_0$. The spectrum is shown as a function of $\omega$ and $k_{\parallel}$ ($k_z$) for fixed $k_x=0$ and $ky=15$. The blue dashed, green dashed-dotted and orange solid lines correspond to the linear dispersion relation of Alfv\'en waves ($\omega_A$), of slow magnetosonic waves ($\omega_F$) and of fast magnetosonic waves ($\omega_S$), respectively. We include the particle gyrofrequency $\omega_g$ and the sweeping frequency in red dashed and blue solid lines, respectively.}
    \label{bx_x_kz}
\end{figure}

For the study of the relevance of waves in the system for different $\tau_c$, we quantitatively analyze the spatio-temporal magnetic spectra.
Figure \ref{bx_x_kz} shows the spatio-temporal spectrum of the perpendicular magnetic field fluctuation component $E_{xx}(k_x=0,k_y=15,k_z,\omega)$ for the NL1, NL2 and NL3 runs.
For an easy comparison, we include the dispersion relation for the Alfvén, fast and slow magnetosonic waves in blue dashed, green dashed-dotted and orange solid lines, respectively.
We also added the particle gyrofrequency and the sweeping frequency in red dashed and solid blue lines, respectively.
We observed that the energy is mainly located around the slow branch and, in lesser extent, around Alfvén waves. Moreover, as $\tau_c$ decreases (i.e., as we move from NL1 to NL3), the energy around wave modes increases for lower values of $k_z$. 
In fact, it is worth noting that as we increase the correlation time the energy tends to spread slightly towards the sweeping frequency.
This result is indeed compatible with the behaviour on $\ell_c/l_d$, as the sweeping energy is mostly related to the non-linear structures (see Table \ref{tab:nl_res}).

\begin{figure}[]
    \centering
    \includegraphics[width=1.0\linewidth]{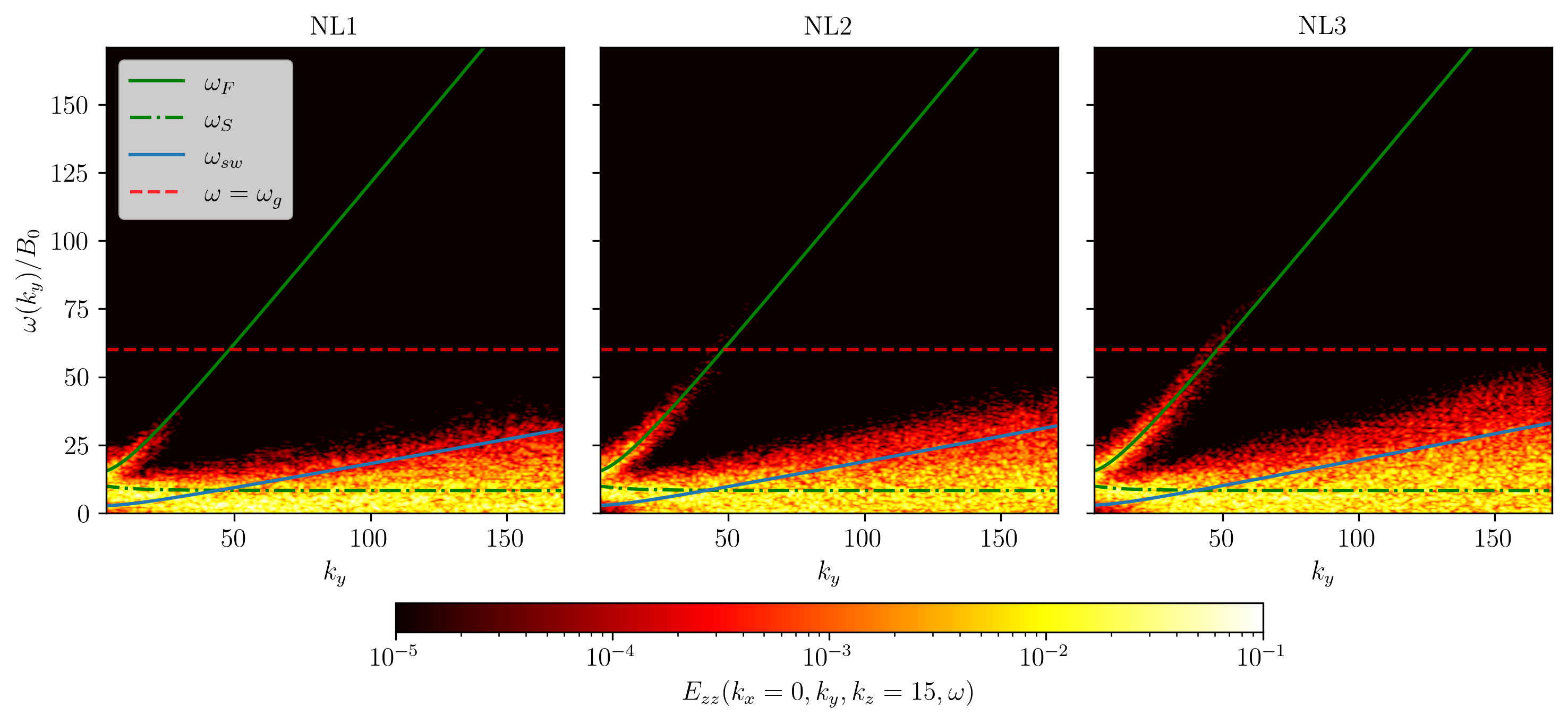}
    \caption{Spatio-temporal spectrum $E_{zz}(k_x=0,k_y,k_z=15)$ for the magnetic field fluctuations parallel to $B_0$, for the runs L and NL1. The spectrum is shown as a function of $\omega$ and $k_{y}$ for fixed $k_x=0$ and $k_{\parallel}=15$. The green solid and the green dashed-dotted lines correspond to the linear dispersion relation of fast magnetosonic waves ($\omega_F$) and of slow magnetosonic waves ($\omega_S$), respectively. We include the particle gyrofrequency $\omega_g$ and the sweeping frequency in red dashed and blue solid lines, respectively.}
    \label{bz_x_ky}
\end{figure}

In order to analyze the magnetosonic modes, we studied the spatio-temporal spectrum of parallel magnetic field fluctuation $E_{zz}(k_x=0,k_y,k_z=15$).
Figure \ref{bz_x_ky} shows the same trend that we observed in Fig.~\ref{bx_x_kz}. In fact, the energy around the fast branch decreases noticeably as we increase $\tau_c$.
It is worth mentioning that the slow branch is completely immerse in the sweeping region, therefore, we can not come to a conclusion about the increase (or decrease) of the magnetic power energy around the slow mode.
However, in the case of NL1, the energy is more concentrated around $\omega = 0$ compared to lower values of the correlation time and the sweeping frequency correctly delimits this energy.

\begin{figure}[]
    \centering
    \includegraphics[width=1.0\linewidth]{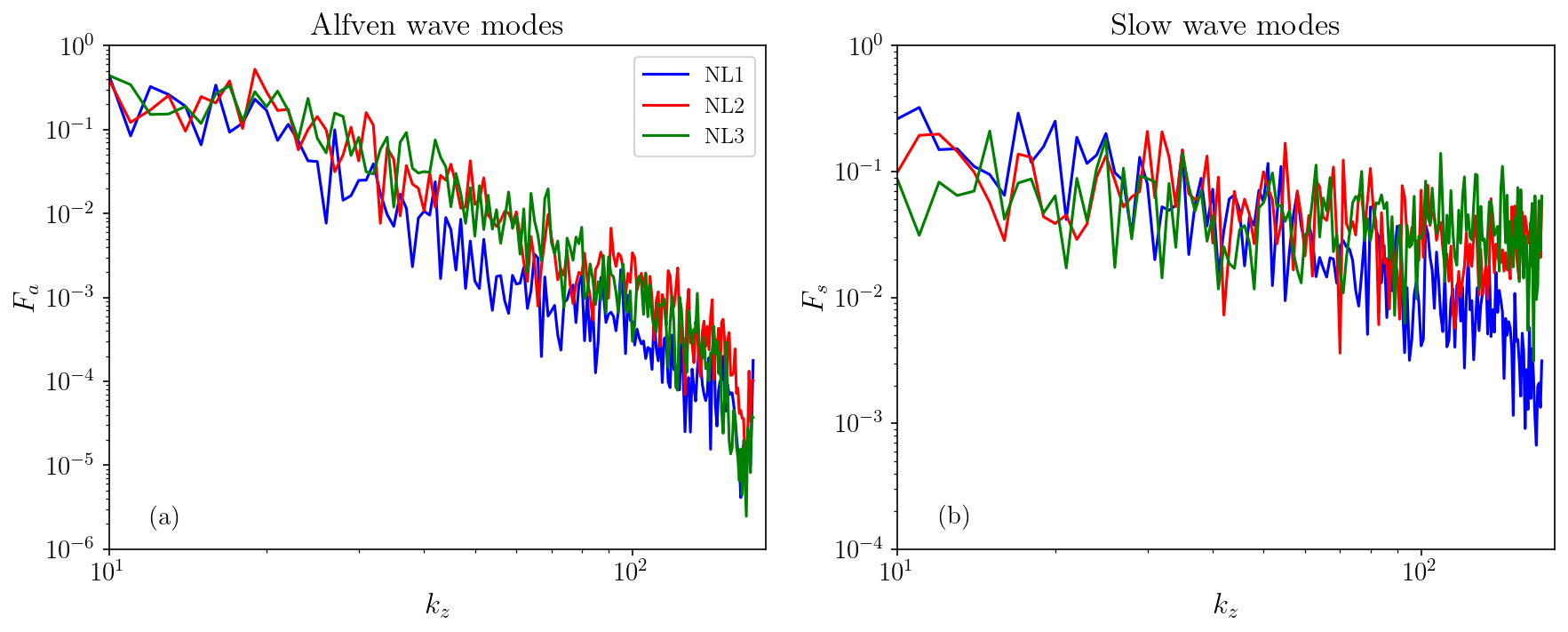}
    \caption{Quantification of the amount of energy around (a) Alfv\'en and (b) slow magnetosonic branches presented in spatio-temporal spectrum $E_{xx}(k_x=0,k_y=15,k_z)$.}
    \label{F_kz}
\end{figure}

\begin{figure}[]
    \centering
    \includegraphics[width=1.0\linewidth]{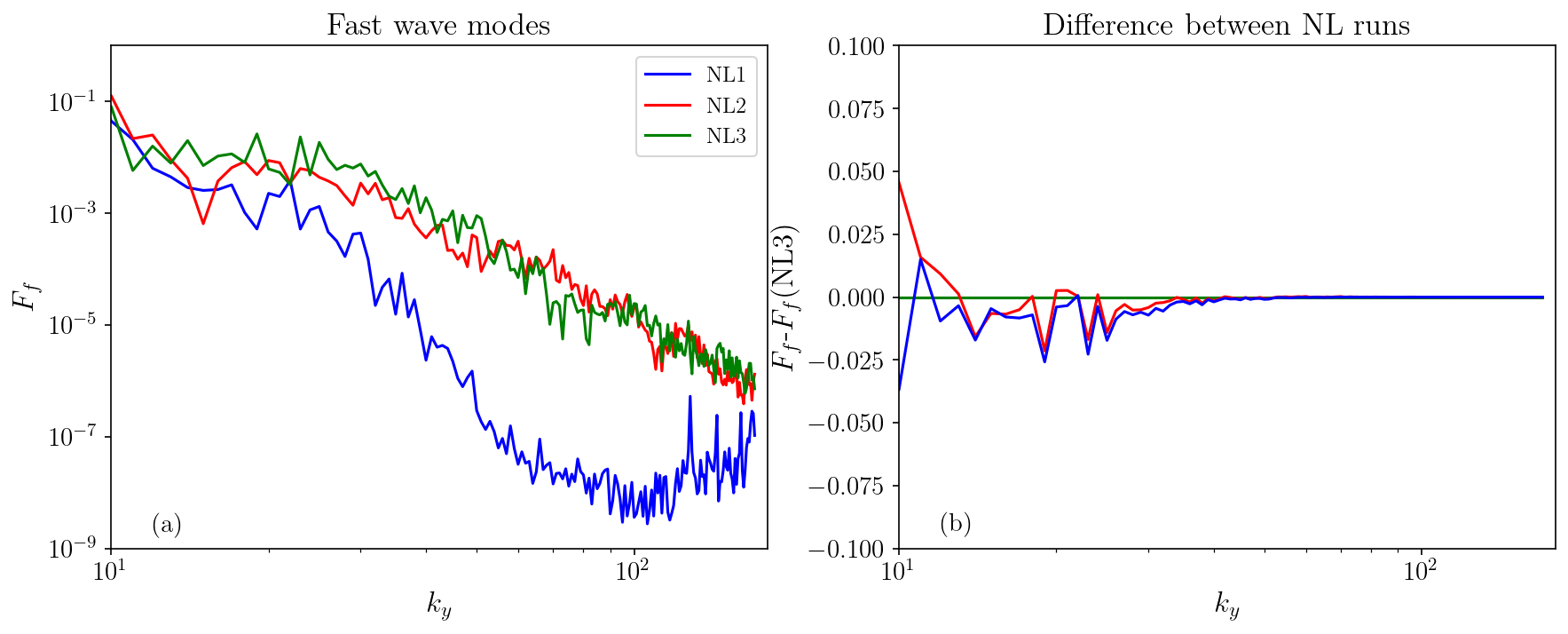}
    \caption{Quantification of the amount of energy around (a) fast magnetosonic branch presented in spatio-temporal spectrum $E_{zz}(k_x=0,k_y,k_z=15)$. Difference between the NL3 run and the rest of the nonlinear simulations (b).}
    \label{F_ky}
\end{figure}

For a more precise study, we used an integration method to quantify the amount of energy near the different wave modes in each spatio-temporal spectrum. \citet{Cl2015} calculated the ratio of the energy accumulated near these wave modes to the total energy in the same wavenumber as,
\begin{equation}
    F(k_z)=\frac{E_{xx}(k_x=0,k_y=0,k_z,\omega=\omega_{A,F,S})}{\Sigma_j E_{xx}(k_x=0,k_y=0,k_z,\omega_j)},
    \label{cuantificacion}
\end{equation}
with $\omega_{A,F,S}$ the frequencies that satisfy a certain dispersion relation (and the $E_{xx}(k_z)$ component is an illustration example only). Figure \ref{F_kz} shows the energy around (a) Alfvén and (b) slow magnetosonic waves for runs NL1, NL2 and NL3.
In particular, we observed that the amount of energy near the slow branch is similar as we increase the correlation time. 
However, for small scales ($k_{z} \sim 100$), the energy in run NL1 tends to be smaller than the rest of the simulations as a result of the energy moving towards the sweeping frequency.
In the case of Alfvén modes, most of the energy is concentrated around low wavenumbers and tends to decrease for higher $k_z$. 
It is worth noting that we are not including the fast branch since most of the energy is not located around this wave mode.
Figure \ref{F_ky} shows (a) the energy near fast magnetosonic waves for runs NL1, NL2 and NL3 and (b) the difference between NL3 and the first two non-linear simulations. 
We observe that the energy located around the fast branch reaches its maximum value for the smallest wavenumbers. 
Moreover, we obtained a significant growth of the energy as the correlation time decreases. 
Finally, the NL1 run differs from the NL3 run by 11$\%$.

\section{Conclusions}\label{sec:disc}

In the present work, we investigated the interplay of linear waves and coherent structures in compressible MHD turbulence and how this affects test particle (proton) energization.
We compared a nonlinear evolution against a linear evolution of the same initial turbulent state.
This initial state is obtained from a fully nonlinear evolution, which is more realistic than the usual spectra models with random phases.
We found that under this initial state, energization in the linear case is much lower than the nonlinear case.
However, this situation inverts after the phases of the initial state are randomized, showing the relevance of phase distribution in particle energization.
This last numerical result allow us to reinterpret the role of phases in particle energization, which are usually treated as secondary with respect to the spectra.
Additionally, this result shows that linear evolution, no matter how realistic the spectra is,  can not faithfully reproduce energization observed from a nonlinear evolution.

We showed that wave-particle interactions are mainly in modes $40 < |\vb{k}| < 80$, which for Alfvén waves is consistent with $\omega_A \approx \omega_g$.
This could be related to second order Fermi energization $\omega \pm k_zv_z=\omega_g$ in the reasonable limit $\omega_g \gg k_zv_z$.
First order Fermi energization requires $|v_z|\approx v_A\approx 9v_0$, which we observed from the parallel energization in Fig. \ref{fig:energy_l_vs_nl} is not likely.
A similar reasoning follows for the fast magnetosonic waves, whose frequency $\omega$ and resonant velocity $\omega/k_z$ are even higher.
On the other hand, first order Fermi energization is more likely to happen in the slow magnetosonic waves and is the only possible mechanism left for the LR40 run. 

The fact that the L and LR40 runs have similar energization mechanisms and rates is striking, as it further reinforces the role of phases in particle energization.
If slow magnetosonic waves are responsible for the energization in the LR40 run, they must also be for that of the L run.
However, this is unlikely, as their phase and energy distributions are different in each run because phase randomization decorrelates and tends to evenly distribute energy between the different branches.
According to our numerical results, it would seem that phase distribution is very important for high wavelength waves and becomes practically irrelevant for low wavelength waves.
This result suggests that the simple picture of particles resonating with one wave at a time may be misguiding and perhaps resonance broadening is enhanced for such complex spectra.
The PDFs of Fig. \ref{fig:2d_vel_hist} seem to imply that the energization of L/LR40 and LR/LR80 are fundamentally different.

Furthermore, we analysed the effect of the correlation time of the external forcings $\tau_c$ on the spatio-temporal spectra and the particle energization.
Here, we further discussed the trapping and energization mechanism of the structures, showing that clustered particles energize exponentially.
Evolutions with higher $\tau_c$ tend to trap particles for longer periods of time, thus allowing them to better exploit this mechanism.
Based on spatio-temporal spectra, we showed that higher values of $\tau_c$ reduce the energy in the fast magnetosonic and Alfvén branches, while also moving energy closer to the sweeping region.
This shows that higher frequency forcings (low $\tau_c$) induce relatively more linear waves in the system, taking away energy from nonlinear structures.
As a result, trapping is less effective because the fluctuations find it easier to remove particles from coherent structures.

Therefore, we could argue that particle energization decreases as the fraction of linear energy increases.
This further confirms that energization by structures is dominant, as the increase in wave energy is not enough to compensate the loss in particle energization due to weaker structures.
We also showed strong evidence of the importance of trapping in the particle energization process and in particular the relationship between this effect and the persistence of structures with the correlation time of the forcing. An interesting follow up could be the study of particles with different charge-to-mass ratio (see previous work in \cite{pugliese2022}). 
Finally, we believe that the present work can shed light on the understanding of the complex process of particle energization in turbulent scenarios \citep{M2023}, like those found in the interplanetary space and more general astrophysical contexts \citep{BC2013}.

\section{Acknowledgment}
\begin{acknowledgments}
The authors acknowledge financial support from CNRS/MINCyT ECOS SUD 2022 No.~A22U02. N.A.~acknowledges financial support from the following grants: UBACyT 20020190200035BA. P.D.~acknowledges financial support from PIP Grant No.~11220150100324
and 11220200101752 and PICT Grant No.~2018-4298. 
\end{acknowledgments}

\bibliography{cites}{}
\bibliographystyle{aasjournal}

\end{document}